\documentclass[aps,prb,reprint,superscriptaddress,amsmath,amssymb]{revtex4-2}

\usepackage{xcolor}
\usepackage{graphicx}
\DeclareGraphicsExtensions{.png,.pdf,.tif,.jpeg}

\begin{document}


\title{Dominance of Electron-Magnon Scattering in Itinerant Ferromagnet Fe$ _{3} $GeTe$ _{2} $}


\author{P. Saha}
\affiliation{School of Physical Sciences, Jawaharlal Nehru University, New Delhi-110067, India}
\author{M. Singh}
\affiliation{School of Physical Sciences, Jawaharlal Nehru University, New Delhi-110067, India}
\author{V. Nagpal}
\affiliation{School of Physical Sciences, Jawaharlal Nehru University, New Delhi-110067, India}
\author{P. Das}
\affiliation{School of Physical Sciences, Jawaharlal Nehru University, New Delhi-110067, India}
\author{S. Patnaik}
\email{spatnaik@jnu.ac.in}
\affiliation{School of Physical Sciences, Jawaharlal Nehru University, New Delhi-110067, India}


\date{\today}

\begin{abstract}
Fe$ _{3} $GeTe$ _{2} $ is a 2-dimensional van der Waals material exhibiting itinerant ferromagnetism upto 230 K. Here, we study aspects of scattering mechanism in Fe$ _{3} $Ge$ _{2} $Te$ _{2} $ single crystals via resistivity, magneto-transport and Hall effect measurements. The quadratic temperature dependence of electrical resistivity below the Curie temperature hints towards the dominance of electron-magnon scattering. A non-saturating positive magnetoresistance (MR) is observed at low temperatures when the magnetic field is applied parallel to the sample plane. The linear negative MR at high fields for $ T<T_{\text{C}} $ corroborates to the suppression in magnon population due to the damping of spin waves. In the high temperature regime $ T>T_{\text{C}} $,MR can be described by the scattering from spin fluctuations using the model described by Khosla and Fischer. Isothermal Hall resistivity curves unveil the presence of anomalous Hall resistivity. Correlation between MR and side jump mechanism further reveals that the electron-magnon scattering is responsible for the side jump contribution to the anomalous Hall effect. Our results provide a clear understanding of the role of electron-magnon scattering on anomalous Hall effect that rules out its origin to be the topological band structure.

\begin{description}
\item[PACS numbers] 01.30.Ww
\item[Keywords] Magnons, Anomalous Hall effect, Magnetoresistance, Spin-fluctuations, Electron-magnon scattering
\end{description}
\end{abstract}


\maketitle


\section{\label{sec:level1}Introduction}
\begin{figure*}[htp]
\includegraphics[width=0.8\textwidth,height=8cm]{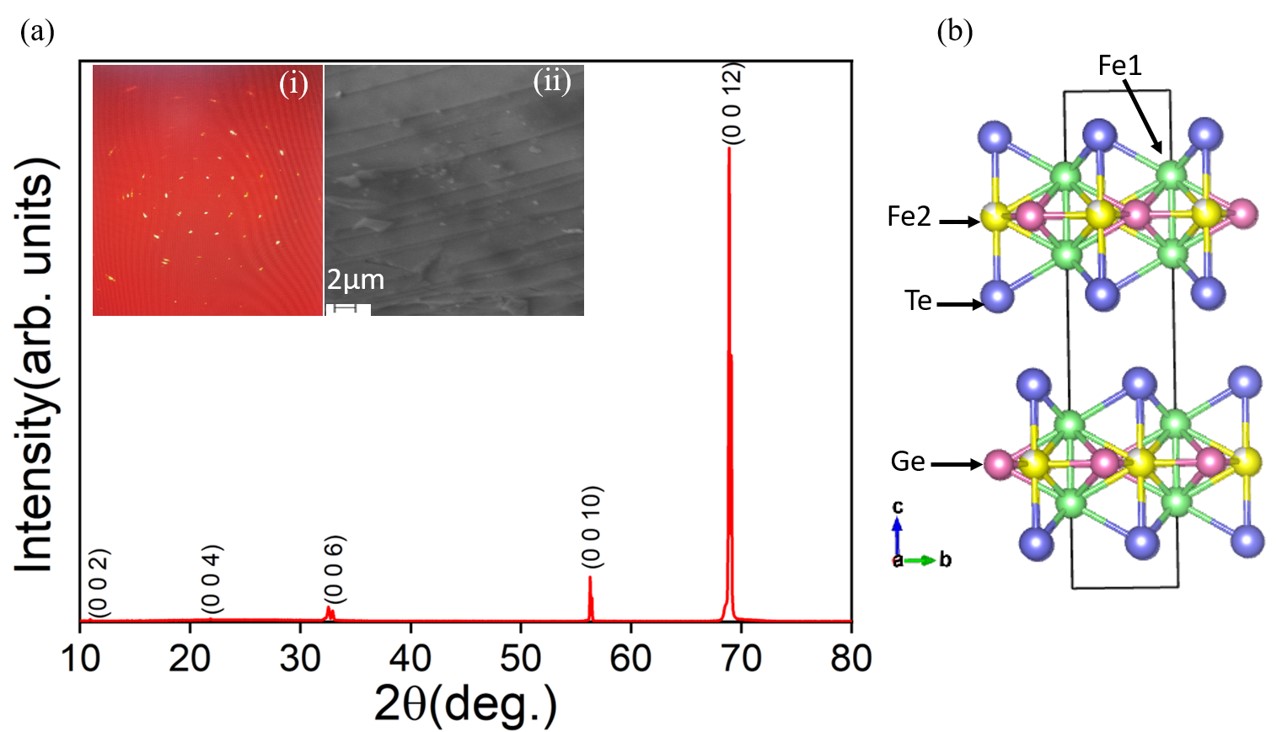} 
  \caption{(a) X-ray diffraction pattern of single crystal FGT. Inset (i) shows the Laue diffraction pattern Inset (ii) SEM image of the crystal depicting the layered structure of FGT single crystal (b) Schematic view of crystal structure of FGT.}
\end{figure*}
Two dimensional (2D) materials have attracted notable attention in the field of electronic devices due to their intriguing physical properties and the feasibility in fabrication of complex structures out of them [1,2]. In recent years,research on 2D materials have been shifted to van der Waals bonded heterostructures for their potential applications [3,4,5]. With the discovery of Cr$ _{2} $Ge$ _{2} $Te$ _{6} $, which is a nearly ideal two-dimensional Heisenberg ferromagnet [6], the magnetism aspects have also come to the fore. In the recent past, Fe$ _{3} $GeTe$ _{2} $(FGT), a 2D van der Waals material has gained significant interest due to its appealing properties such as uniaxial magnetocrystalline anisotropy[7], Kondo lattice behaviour[8], large anomalous Hall current[9] and ionic gate tunable room temperature ferromagnetism[10]. Although the first report of synthesis of FGT was published in 2006 [11], its high temperature itinerant ferromagnetism (below 230 K) was reported recently [12]. Further, the magnetic force microscopy (MFM) measurements and density functional calculations have confirmed that this compound has an additional antiferromagnetic ground state below 152 K due to the oppositely aligned spins of Fe atoms between the adjacent layers[13]. It was proposed to host two competing magnetic orderings between 152 K and 214K [13]. Previous reports have suggested that the ferromagnetic transition temperature $ T_{\text{C}} $ and the lattice parameters can be tuned by controlling the concentration of the Fe [14]. Hall effect measurements on FGT have revealed conventional anomalous behaviour when field is applied along the easy axis and it showed signatures of topological Hall effect when the field was applied along the ab plane[15,16]. 
Magneto-transport properties of low $ T_{\text{C}} \sim $138 K phase of FGT have been extensively studied [17]. Here, we report synthesis of FGT with a much higher $ T_{\text{C}} \sim $~210 K and have focussed to unravel the scattering mechanisms responsible for the magneto-transport behaviour in FGT and its role in the temperature dependence of anomalous Hall effect (AHE). 
\par In general, three different scattering mechanisms need to be taken into account to explain the origin of the AHE. One is the extrinsic skew scattering mechanism [18,19] which arises due to the intertwinement of scattering potentials and spin orbit coupling. It has a linear dependence on longitudinal resistivity($ \rho_{\text{xx}} $). Second is the extrinsic side jump mechanism[20] which arises due to the transverse shift experienced by the charge carriers due to the scattering in the presence of spin-orbit interaction which leads to a quadratic dependence to longitudinal resistivity. Third, is the intrinsic Karplus Luttinger (KL) mechanism [21] which results in an ‘anomalous velocity’ originating from the Berry curvature of the occupied eigenstates and is related to the band structure of the respective material. It also has a quadratic dependence to longitudinal resistivity. Nevertheless, the temperature-dependent change of AHE remains vague and open to debate in both theoretical and experimental studies. The contribution to AHE due to KL mechanism is an inherent ground state feature which is not supposed to change with temperature[21]. However, it has been demonstrated experimentally, that the extrinsic side jump contribution to AHE can vary as a fuction of temperature[22]. Side jump contribution not only includes the coordinate shift of the wave packet[20] but also the scattering induced contributions due to phonons and magnons[23]. Yang et. al. have proposed that magnons tend to play a distinct role in the AHE as compared to phonons and other impurities[23]. Moreover, the electron-magnon scattering can control the temperature dependence of side jump mechanism[24].
\par Being a 2D van der Waals material, an itinerant ferromagnet with high $ T_{\text{C}} $ and a candidate of nodal line semimetal[9], FGT has been the best platform to study the interrelation between ferromagnetism and topology[9]. It exhibits a substantially large anomalous Hall current(AHC), which stems from the large Berry curvature associated with the nodal line. There are suggestions that AHE in FGT may be explained by the KL mechanism[9,15,16]. However it still poses open questions regarding temperature dependence of AHE and MR and its association with specific scattering mechanism. In this paper, the microscopic origin of scattering mechanisms which affect the magnetoresistive property of FGT has been discussed. In order to study the temperature dependence of AHE, the magnetoresistive behaviour has been correlated to the side jump contribution. Our results suggest that magnons play an important role in the MR and temperature dependence of AHE in FGT. This makes FGT a perfect testing ground for theoretical interpretations. 
 
\section{\label{sec:level2}Experimental Techniques}
The single crystals of Fe$ _{3} $GeTe$ _{2} $ were synthesized by chemical vapour transport (CVT) method. Powders of Iron (99.9\%), Germanium(99.999\%) and Tellurium(99.9\%) were taken in the stoichiometric ratio and were ground for half an hour using mortar and pestle. The homogeneous mixture was then cold pressed into pellets. These pellets were then inserted in 30 cm long quartz tube with iodine (2mg/cm$ ^{3} $) as the transport agent. The tube was vacuum sealed and placed in a two-zone furnace with a temperature gradient of 750-700 $ \deg $C for a week. The plate shaped crystals were deposited in the low temperature zone of the tube. The crystal structure and phase at room temperature was identified using X-ray diffraction(XRD) in Rigaku Miniflex 600 instrument . Bruker X-ray diffractometer was used to perform the single crystal XRD. The magnetotransport measurements were performed using a Cryogenic built Cryogen Free Magnet (CFM) (8T,1.6K) and the temperature dependent magnetization measurements were performed using the Vibrating Sample Magnetometer(VSM) attachment of a Physical Properties Measurement System (PPMS). Scanning Electron Microscopy (SEM) was carried out using a Zeiss EVO40 SEM analyser. 	
\begin{figure*}
\includegraphics[width=0.8\textwidth,height=12cm]{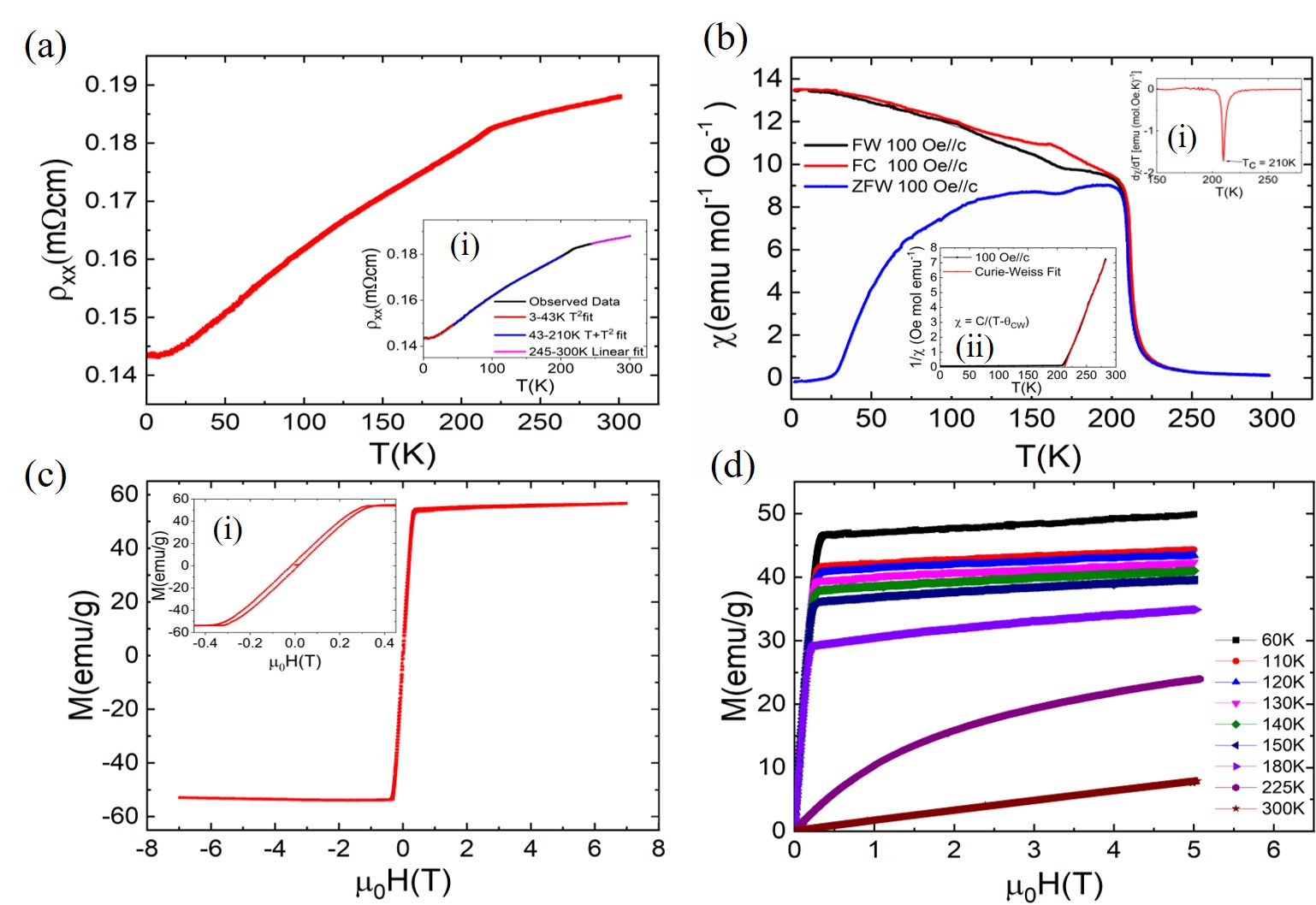} 
  \caption{(a) Temperature dependence of longitudinal resistivity. Inset(i) in (a) shows the theoretical fits for $\rho$(T) for different temperature regimes. (b) is the temperature dependent susceptibility for zero field cooled, field cooled and field warming at $ H=100 $ Oe. Inset (i) in (b) shows $d\chi/dT$ vs $ T $ behaviour. Minima of the plot represents the Curie temperature at $T_{\text{C}}=210$ K. Inset (ii) in (b) shows the temperature dependence of inverse susceptibility 1/$\chi$ and it’s fitting using the Curie Weiss equation.(c) shows the zoomed out view of the magnetization data at 2 K upto ±$7$ T. Inset (i) of (c) shows the magnified view of field dependent magnetization behaviour at 2 K which shows a clear hysteresis. (d) Isothermal magnetization data at different temperatures upto 5 T for $ H \parallel c $.}
\end{figure*}
\section{\label{sec:level3}Results and Discussions}
\subsection{X-ray Diffraction}
Fig.1(a) shows the XRD diffraction pattern of a single crystal flake of FGT. The diffraction peaks dominantly correspond to the $00l$ planes and thus it depicts the single crystalline nature of as-grown samples. Further, the Laue spots were verified using single crystal diffractometer (inset (i) of Fig 1.(a)).  The lattice parameters are obtained as $ a = b = 4.01$ \AA{} and $ c = 16.46 $\AA{} which is in good agreement with hexagonal crystal structure (space group $P63/mmc$) [11]. The SEM image of FGT crystal is shown in fig.1(a) inset(ii) that reflects the layered structure of the sample. The unit cell schematic of FGT is shown in fig.1(b). There are two inequivalent sites for Fe; Fe1 (green) forms a hexagonal arrangement with only Fe atoms and Fe2 (yellow) is covalently bonded to germanium in adjacent layer. FGT has a structure in which the covalently bonded Fe$ _{3} $Ge slabs are sandwiched by layers of tellurium atoms. Adjacent layers of tellurium atoms are bonded through van der Waals interaction.
\subsection{Resistivity}
Fig. 2(a) shows the temperature dependent resistivity $ \rho $ behaviour of FGT sample upto the room temperature. The resistivity increases with temperature implying the metallic nature. There is an abrupt change in the slope at 215 K which is reflective of onset of a temperature driven ferromagnetic phase transition. It is observed that residual resistivity (0.14 m$ \Omega $-cm) is lower than earlier reports [9,15]. 
In general, the resistivity of a metallic sample is determined by several scattering mechanisms. In the following, $ \rho $ vs $ T $ behaviour is studied in FGT in different temperature ranges and corresponding microscopic mechanism has been identified. It is found that the temperature dependence of resistivity changes its nature in specific ranges; 3 K-43 K, 43 K-210 K and 245 K-300 K.  In the ferromagnetic state, electron-magnon scattering may contribute significantly. This is derived from the quadratic dependence of resistivity with temperature $ (\rho \varpropto T^{2}) $. Inset of Fig.2(a) shows the observed data fitted with the theoretical curves for the different temperature ranges. In the low temperature region (3 K $ < T < $ 43 K), the temperature dependent resistivity shows a complete quadratic behaviour $ (\rho \varpropto T^{2}) $ without any contribution from lower exponents. The behaviour is ascribed to the electron-magnon scattering. As the temperature increases, that is in the range 43 K $ < T < $ 210 K, the resistivity data follows a sum of linear and quadratic dependence with temperature. This is evidence for admixture of electron-phonon scattering  along with electron-magnon scattering. For temperatures above 245 K,which is a paramagnetic phase a complete linear behaviour has been obtained $ (\rho \varpropto T) $. This is ascribed to the dominance of electron-phonon scattering mechanism.

\subsection{Magnetization}

The main panel of Fig.2(b) shows the temperature dependent magnetic susceptibility $ \chi $ curve in the presence of magnetic field $ H = 100 $ Oe applied perpendicular to the ab plane of the sample under zero field cooled (ZFC), field cooled cooling (FCC) and field cooled warming (FCW) protocols. A steep growth in FC and ZFC susceptibility curves are seen at $ T = 210 $ K. This clearly marks the magnetic phase transition taking place in the compound. The Curie transition temperature is determined to be 210 K through the minimum in $ d\chi/dT $ vs $  T $ curve (fig.2(b) inset (i)). Just below the Curie temperature a notable splitting between FCW, FCC and ZFC is observed. Similar behaviour has been reported previously in FGT and has been attributed to the presence of irreversibility of ferromagnetic domains [11]. All the three curves show a kink around 165 K which indicate the presence of a different magnetic phase identified as the competing ferromagnetic and antiferromagnetic phases [13]. However, some reports demonstrated that this dip in the three curves is due to the formation of Neel type chiral spin spirals [25]. An additional kink is observed in ZFC data at 30 K, where the susceptibility drops to zero. This result confirms the antiferromagnetic nature of the sample at lower temperatures and is consistent with the previous reports [13]. Fig. 2(b) inset(ii) shows the temperature dependence of inverse magnetic susceptibility $ 1/\chi $. The paramagnetic region of the data ($ T > $ 210 K) obeys the Curie Weiss law.
 \begin{equation}
\chi = \frac{C}{T-\theta}                                                                                    \end{equation}                                                                                   
where $ C $  is the Curie constant and $ \theta $ is the Curie Weiss temperature. The obtained values through this fit are $ C = 9.84 $ emu K/mol and $ \theta = 212 $ K. This positive value of Curie Weiss temperature confirms the dominant ferromagnetic exchange interactions. The Curie constant is related to the number of unpaired electrons in the sample and the effective moment per magnetic ion is calculated to be $ 5.12\mu_{\text{B}}/Fe $ which is near to the theoretical value of $ 4.90\mu_{\text{B}}/Fe $ for Fe$ ^{2+} $ ion. Fig. 2(c) shows the magnetic field dependent magnetization $ M-H $ curve at 2 K. Magnetization saturates to 50 emu/g at 0.36 T. This saturation depicts the ferromagnetic behaviour and a clear hysteresis is observed in Fig.2(c) inset with coercive field $ H_{\text{C}}= $ 0.16 T. Fig. 2(d) shows the isothermal magnetization curves at different temperatures. The steep increment in magnetization followed by a saturation further depicts the robust ferromagnetic nature of FGT. 
\begin{figure*}[htp]
\includegraphics[width=\textwidth,height=9cm]{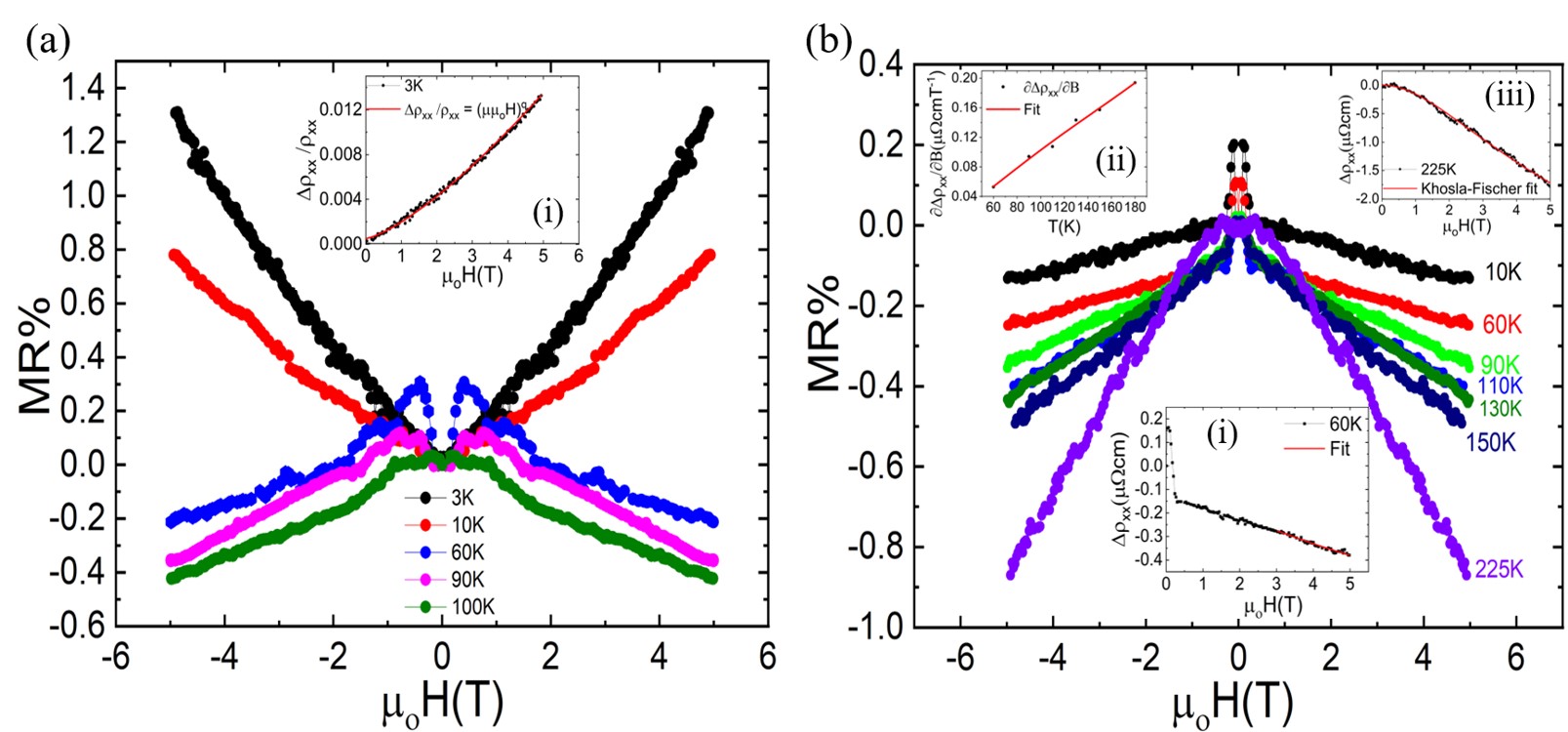}
\caption{(a) shows MR at different temperatures for $ H \parallel ab $. Inset of 3(a) shows the field dependent $\Delta\rho_{xx}$/$\rho_{xx}$ at 3 K and its corresponding fit using equation(6). (b) shows MR at different temperatures for $ H \parallel c $. Inset (i) of (b) field dependent change in longitudinal resistivity $\Delta\rho_{xx}$ at 60 K and the red line is fit to equation(7). Inset (ii) of (b) shows the variation of $\partial\Delta\rho_{xx}$/$\partial B$  with temperature. The red line is the fit using equation(8). Inset (iii) of (b) shows the field dependent behaviour of $\Delta\rho_{xx}$ at 225 K with it’s Khosla and Fischer fit (equation(9)) represented by the red line.}
\end{figure*}

\subsection{Magnetoresistance}

The resistivity is customarily defined as a function of the relaxation time as defined below [26]
\begin{equation}
    \rho_{total}=a_{1}(\omega_{\text{c}}\tau)^n+a_{2}(1/\tau)
\end{equation}
 where $ \tau $ is the electronic relaxation time and $\omega_{c}$  is the cyclotron frequency.
The first term originates from the Lorentz force which restricts the motion of the free carriers. This constrained motion of the carriers raises the electrical resistivity with the increasing field and hence results in a positive MR. The second term is a sum of the contribution of the different scattering processes which tends to obstruct the conductivity and can be expressed as per the Matthiessen’s rule:
\begin{equation}
    a_{2}(1/\tau)= \rho_{\text{res}}+\rho_{\text{e-e}}(T)+\rho_{\text{ph}}(T)+\rho_{\text{mag}}(T)
\end{equation}
Here the first term $ \rho_{\text{res}} $  is due to the scattering due to impurities. Further, $ \rho_{\text{e-e}}, \rho_{\text{ph}} $ and $ \rho_{\text{mag}} $ is due to electron-electron, electron-phonon and electron-magnon scattering, respectively. The first three contributions (impurities, electrons and phonons) are weakly dependent on external magnetic field whereas the electron magnon scattering is a field dependent term. Magnons are the quasiparticles associated with the collective excitations in the spin ordered ground state. At low temperatures, the probability of spin flip transition reduces and so the population of magnons decline. Hence, the electron-magnon scattering is much enhanced at higher temperatures as compared to lower temperatures. However, the application of magnetic field tends to dampen these spin waves and this leads to suppression of the electron magnon scattering. As a result, the longitudinal resistivity decreases with increasing magnetic field that results in a negative MR.
\newline MR measurements on FGT crystals were performed with magnetic field upto 5 T applied perpendicular and parallel to the sample plane. In order to remove the contribution from Hall resistivity from the longitudinal MR the following equation was used:
\begin{equation}
    \rho_{\text{xx}}(H,T)=\frac{\rho_{\text{xx}}(H,T)+\rho_{\text{xx}}(-H,T)}{2}
\end{equation}
Fig.3(a) and Fig.3(b)shows the isothermal MR with field applied parallel and perpendicular to the c axis. The MR is calculated using the following equation:
\begin{equation}
    \frac{\Delta\rho_{\text{xx}}}{\rho_{\text{xx}}(T,0)}=\frac{\rho_{\text{xx}}(T,H)-\rho_{\text{xx}}(T,0)}{\rho_{\text{xx}}(T,0)}
\end{equation}                                                                                     
Where $ \rho_{\text{xx}}(T,0) $  and $ \rho_{\text{xx}}(T,H) $  are the longitudinal resistivities at zero and nonzero field. For $ H \parallel ab $, a positive MR at low temperature is observed. The sign of MR shifts from positive to negative at higher temperatures Fig.3(a). This crossover behaviour of MR has not been observed in bulk FGT thus far. For $ T> $10 K under $ H \parallel ab $ and for all temperatures in $ H \parallel c $ a weak upturn in resistivity is observed which crosses into a negatively linear MR. This upturn in resistivity is obtained at fields for which $ M<M_{\text{s}} $. This behaviour is attributed to the scattering at domain walls which tend to enhance the resistivity. With the increasing field, M reaches Ms and the domain walls are annihilated after which only a negatively linear MR is observed. However, for higher temperature region($T> T_{\text{C}} $) fig.3(b), in the paramagnetic phase of FGT, due to the absence of domain walls, there is a conspicuous change in the behaviour of the curve.
At low temperatures (for $ H \parallel ab $), the positive MR indicates the dominance of the orbital MR (first term of equation (1) where \(\displaystyle\rho_{xx}\propto(\omega_{c}\tau)^n\). The field dependent change in resistivity is given by 
 \begin{equation}
 \frac{\Delta\rho_{\text{xx}}}{\rho_{\text{xx}}(0)}=(\mu\mu_{0}H)^q
 \end{equation}
Where is $ \mu $ the mobility and the exponent $ q = 2 $ according to standard theories [27]. However, experimental results have shown deviation from  $ q = 2 $ with $ 1< q <2 $ has been noted in many systems including ferromagnetic thin films[28,29] and doped semiconductors[30].The inset of fig 3b. shows the low temperature MR at 3 K fitted using the equation(6). The parameters obtained are  $ \mu = 0.0024 $  m$ ^{2} $/V-s and  $ q = 1.34 $ for 3 K and $ \mu = 0.0016 $ m$ ^{2} $/V-s and $ q = 1.27 $ for 10 K.
At moderate temperatures ($ H \parallel ab $) and for $ H \parallel c $, a linearly non-saturating negative MR at high fields is observed. The negative MR\% increases with increasing temperature. This behaviour is a clear indication towards the dominance of the electron-magnon scattering in this temperature range and suppression of the same with evolving field. The amount of magnons is high at higher temperatures and the application of high field suppresses the amount of magnons and this results in larger negative MR at high temperatures. Raquet et al[29] have provided an equation to describe the negative MR due to the electron-magnon scattering which is valid for fields below 100 T and for the temperature range of $ T_{\text{C}}/5 $ to $ T_{\text{C}}/2 $ .
\begin{equation}
    \Delta\rho_{\text{xx}}(T,B) \propto \frac{BT}{D(T)^2}\ln\frac{\mu_{\text{B}}B}{k_{\text{B}}T}
\end{equation} 
\begin{figure*}[htp]
\includegraphics[width=0.9\textwidth,height=12cm]{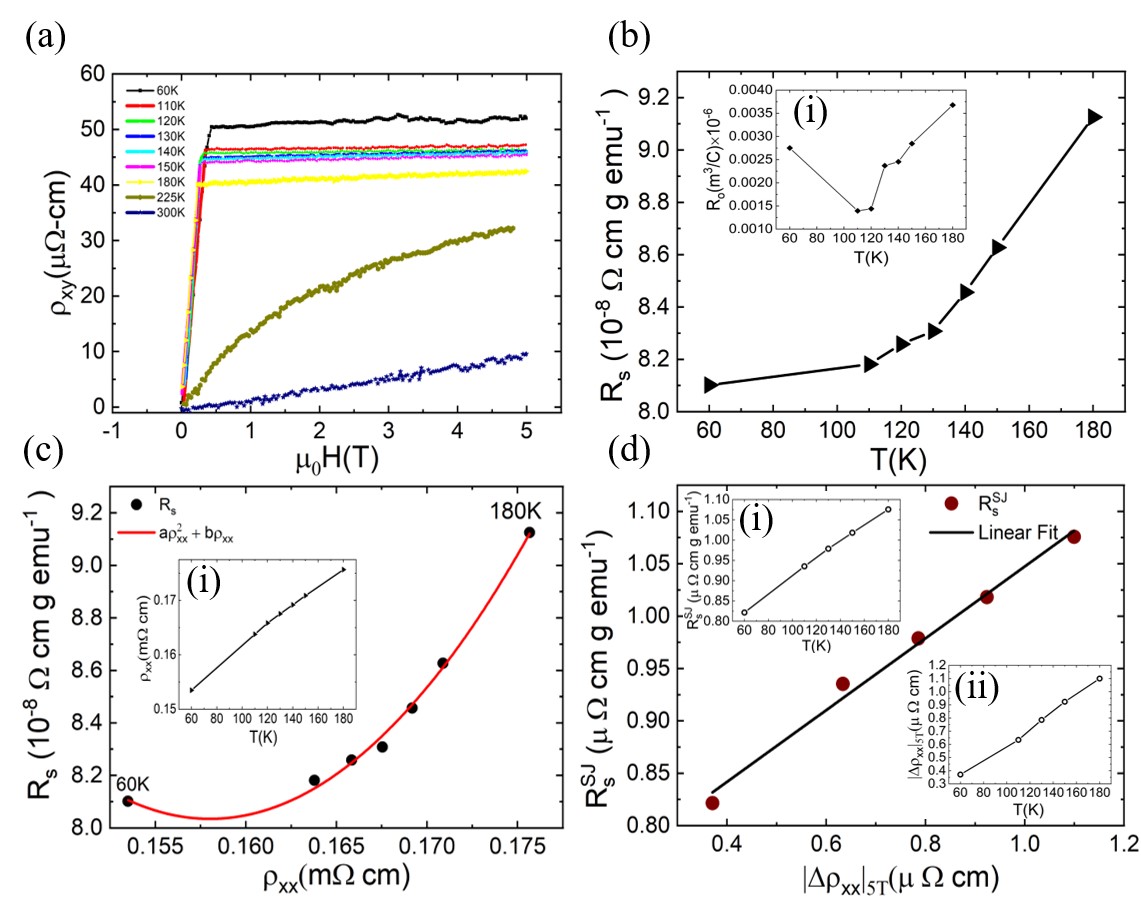} 
  \caption{(a) Isothermal hall resistivity curves as a function of magnetic field at distinct temperatures. (b) Temperature dependence of anomalous Hall coefficient $R_{s}$ determined using equation(11). Inset of (b) shows the variation of normal hall coefficient $R_{o}$ with temperature. (c) scaling of $R_{s}$ vs $\rho_{xx}$ according to equation(12). Inset of (c) shows variation of $\rho_{xx}$ with temperature. (d) shows linear fit between side jump contribution to anomalous hall coefficient, $R_{sj}$ and change in resistivity with field $(|\Delta\rho_{xx}|_{5T})$. Inset (i) and (ii) shows the variation of $R_{sj}$ and $(|\Delta\rho_{xx}|_{5T})$ as a function of temperature.}
\end{figure*}
here $ D(T) $ is the magnon stiffness or the magnon mass renormalization, $ B $ is the magnetic field and $ T $ is the temperature. The first order approximation of magnon stiffness is described as $ D(T)\sim D_{0}(1-d_{1}T^{2}) $ where $ D_{0} $ is the zero temperature magnon mass and $ d_{1} $ is a constant [29,31].The inset(i) of fig.3(b) shows that the field dependent longitudinal resistivity data fits well with the above equation and this confirms that the suppression in electron-magnon scattering is responsible for the linearly negative MR. Furthermore, the slope of the high field MR showed a significant dependence with temperature and is described by the following equation [29]:
\begin{equation}
    \frac{\partial\Delta\rho_{\text{xx}}}{\partial B} \propto T(1+2d_{1}T^2)(\ln(T)+c_{\text{te}})
\end{equation}
Where $d_{1}$ is a constant which depends on the zero temperature magnon mass and $c_{te}$ is a temperature independent term. $d_{1}$ is of the order of $ 10^{-6} $ K$ ^{-2} $ for Fe, Co and Ni thin films[29]. The temperature dependent variation of high field resistivity slope is shown in the inset (ii) of fig.3(b). The above equation provides a good fit for the data with $d_{1} =10^{-5}$ K$^{-2}$. The large value of $d_{1}$ depicts a stronger dependence of magnon stiffness on temperature.
For $ T>T_{\text{C}} $, an enhancement in the negative MR\% is observed. This arises due to the scattering of conduction electrons from the fluctuating local moments. In this regime, a negative nonlinear MR is observed which is unlike the linear MR for $ T<T_{\text{C}}$. Khosla and Fischer[30] have proposed a model to study the scattering from In impurities in CdS and have subsequently predicted this kind of MR. It is described by the following equation:
\begin{equation}
    \Delta\rho_{\text{xx}}=-b_{1}\ln[1+(b_{2}\mu_{0}H)^2]
\end{equation}                                        
Where $ b_{1} $ and $ b_{2} $ are constants. For $ T>T_{\text{C}} $ , spin fluctuations become predominant and the MR is best described by the semiempirical formula of Khosla and Fischer as shown in inset (iii) of Fig.3(b).The parameters returned from this fit are $ b_{1}= 1.08 \mu\Omega $cm and $ b_{2}= 0.39 $  m$ ^{2} $/V-s. It is noteworthy that the Khosla-Fischer formula does not fit for the isothermal field dependent resistivity in the range $T< T_{\text{C}} $ and this marks towards the magnetic phase transition and independently confirms that different scattering processes are involved in these two regimes.

\subsection{Anomalous Hall Effect}

In ferromagnets, the magnetic field driven evolution of the transverse resistivity  $\rho_{\text{xy}}$ is known as the anomalous Hall resistivity ($\rho_{\text{xy}}^{\text{ah}}$). It is described by the following equation:
\begin{equation}
    \rho_{\text{xy}}=\rho_{\text{xy}}^{\text{oh}}+\rho_{\text{xy}}^{\text{ah}}= R_{\text{o}}\mu_{\text{o}}H+4\pi R_{\text{s}}M_{\text{s}}
\end{equation}          
Where $ R_{\text{o}} $ and $ R_{\text{s}} $  are the ordinary and anomalous Hall coefficients respectively. $ M_{\text{s}} $ is the saturation magnetization obtained from the isothermal magnetization curves as shown in fig.2(d). The first term is called the ordinary Hall resistivity which arises due to the deflection of charge carriers as a consequence of the Lorentz force acting on them. The second term is the anomalous Hall resistivity. Due to this additional term, the total Hall resistivity experiences a sharp rise with the evolving field followed by a saturation, much like the field dependent magnetization behaviour. 
Hall resistivity is the transverse resistivity measured at constant temperature with evolving magnetic field as shown in fig.4(a). For temperatures below $ T_{\text{C}} $, there is a steep rise in the Hall resistivity upto a particular field after which it almost reaches saturation. This is a clear indication of the presence of a nonzero anomalous Hall response in our sample.When field was applied along the ab plane, the Hall resistivity shows a cusp like irregularity.This is due to the gauge field which stems from the non-coplanar spin configuration and is congruent with the previous reports[15,16]. The normal Hall coefficient $ R_{\text{o}} $ can be determined by the slope of the Hall resistivity data in the high field region. The sign of this slope determines the type of charge carriers involved. Inset(i) of fig.4(b)shows that $ R_{\text{o}} $ tends to increase with temperature. It’s positive sign at all temperatures indicates that the majority charge carriers are holes. The anomalous Hall resistivity $\rho_{\text{xy}}^{\text{ah}}$ is obtained by extrapolating the Hall resistivity data from the high field to the y axis. The anomalous Hall coefficient is determined using the following relation:
\begin{equation}
    R_{s}=\frac{\rho_{xy}^{ah}}{4\pi M_{S}}
\end{equation}                                                                                                   Fig.4(b) shows $ R_{\text{s}} $ that increases considerably with temperature. It is approximately three orders of magnitude larger than $ R_{\text{o}} $.This shows the strong dominance of the anomalous Hall resistivity over the total Hall resistivity. Lorentz force deflection cannot interpret this large $ R_{\text{s}} $. In ferromagnets the $ R_{\text{s}} $ is a function of longitudinal resistivity $\rho_{\text{xx}}$ specified by the following relation:
\begin{equation}
   R_{s}=a\rho_{xx}^2+b\rho_{xx}
\end{equation}
Here, $ a $ denotes the strength of the side jump contribution[20] as well as the intrinsic Berry phase contribution[21] and $ b $ corresponds to the strength of the skew scattering contribution[18,19].In order to determine the dominant scattering contribution, $R_{\text{s}}(T)$ has been scaled with $\rho_{xx}$ using equation (12) as shown in fig.4(c).The parameters returned from the fit are $a = 34.86$g emu$^{-1} \Omega^{-1}$ cm$^{-1}$ and $ b= -0.011  $ gemu$ ^{-1} $. Inset of fig. 4(c) shows the temperature dependent change of $\rho_{xx}$. It is evident from the parameters that the intrinsic Berry phase and/or side jump contribution ($R_{s}^{sj,i}$) dominates the anomalous Hall effect. This result is in agreement with the earlier reports[9,15]. However, separating the extrinsic side jump ($R_{s}^{sj}$) and intrinsic Berry phase related contribution ($R_{s}^{i}$) is a challenge, since both of them show quadratic dependence to longitudinal resistivity $(\propto \rho_{xx}^2)$. The negative sign of the skew scattering contribution $ b $, indicates that it is acting in the opposite direction as compared to $R_{s}^{sj,i}$.
The intrinsic contribution to anomalous Hall effect is weakly dependent on temperature even though $\rho_{xx}$ varies with temperature[33]. It has already been demonstrated that the electron-magnon scattering can affect the side jump scattering[24] as well as the Berry phase[30]. In FGT, both the temperature dependent resistivity and MR measurement results have shown the dominance of the electron-magnon scattering. In order to confirm the role of magnons and its affect on the temperature dependence of side jump, the temperature dependence change in resistivity under field $|\Delta\rho_{xx}|_{5T}$ is plotted with and $R_{s}^{sj}$. Inset (i) and (ii) of fig. 4(d) shows the change in $R_{s}^{sj}$ and $|\Delta\rho_{xx}|_{5T}$    with temperature.  It is evident from the linear fit of the data in fig.4(d) that the temperature dependence of $R_{s}^{sj}$ stems from the electron-magnon scattering.

\section{\label{sec:level2} CONCLUSION:}
In summary, analysis of temperature dependent resistivity and MR have revealed the dominant scattering mechanisms in different temperature regimes in single crystals of Fe$ _{3} $GeTe$ _{2} $. The large MR at high temperature is explained by the scattering from spin fluctuations. At moderate temperatures, the negative MR at high field is described by the reduction in electron-magnon scattering. This effect tends to weaken as the temperature reduces. For low temperatures, when the field is applied parallel to the sample plane,a positive MR is obtained. This implies dominance of the Lorentz force on the charge carriers. Below $ T_{\text{C}} $, the anomalous Hall coefficient scales with the longitudinal resistivity and reveals that the AHE is driven by the intrinsic Berry phase mechanism and/or side jump mechanism. A one to one mapping between the MR and side jump contribution confirms that the side jump contribution originates from the electron-magnon scattering. In essence, the temperature dependence of AHE in FGT is adequately described by extrinsic sources without invoking topological band structure.

\section*{Acknowledgement}
P.Saha, V. Nagpal, P.Das acknowledge UGC-NET JRF for financial support. M.Singh thanks CSIR for providing JRF. We are grateful to FIST program of Department of Science and Technology, Government of India for low temperature high magnetic field measurement facility at JNU. We are thankful to Advanced Instrumentation Research Facility (AIRF), JNU for PPMS measurement facility.
\nocite{*}
\bibliography{FGT.bib}

\end{document}